# Five-dimensional bigravity.
# New topological description of the Universe.

**Jean-Pierre Petit and Gilles d'Agostini**

**jppetit1937@yahoo.fr**

_______________________________________________________________________

**Abstract:**

We extend the bimetric description of the Universe to a five-dimensional framework. Starting from Souriau's work (1964) we use Robertson-Walker metrics, with an extra term corresponding to the additional Kaluza dimension. This first order model is limited to zero densities of electric charge and electromagnetic energy. Assuming that the massive particles, with positive or negative mass and energies have finite lifetimes, it restores the O(3) symmetry and makes the generalized gauge process to restart. As a consequence the speed of light tends to zero.

We assume the Universe to be closed in all its dimensions. Then, following an idea introduced in 1994 we describe the Universe as the two folds cover of a projective space. The inversion of the arrow of time, mass and energy is a consequence of this geometrical hypothesis and fits the bimetric model. We choose to eliminate the "initial singularity" replaced by a boundary space, which is found to be Euclidean.

_______________________________________________________________________

## 1) Introduction

The theory of the General Relativity was extended to a five-dimensional framework in 1964 by the French mathematician Jean-Marie Souriau in the chapter five of his book "Géométrie et relativité" [1]. He assumes the Universe to be a five dimensional manifold M5. Page 388 he defines his notations and we will choose the same. When the indexes are labelled by Latin letters j , k , l , m,  it means that the refer to five values { 1 , 2 , 3 , 4 , 5 }. When Greek letters are used they refer to a set of four { 1 , 2 , 3 , 4 }. He introduces a Lagragian density, that he calls "presence":

(1)

$$p = a + b\, g_{lm}\, R^{lm}$$

or, in a simpler way:

(2)

$$p = a + b\, R$$

where $R$ is the scalar curvature. The fifth dimension is space-like, whence the signature of the metric is

(3)

$$( + - - - - )$$

If not, the zero divergence condition would not give the classical equations of physics, in non-relativistic Newtonian approximation. Following the Jordan-Thiry approximation Souriau assumes the $g_{jk}$ of the 5d metric do not depend on the fifth variable $x^5$:

(4)

$$\partial_5 \, g_{jk} = 0$$

and writes t:

(5)

$$x = \begin{bmatrix} x^1 \\ x^2 \\ x^3 \\ x^4 \\ x^5 \end{bmatrix} \qquad \hat{x} = \begin{bmatrix} x^1 \\ x^2 \\ x^3 \\ x^5 \end{bmatrix} \qquad x = \begin{bmatrix} \hat{x} \\ x^5 \end{bmatrix}$$

The Universe U is a bundle. The fifth dimension is supposed to be closed, then Souriau writes:

(6)

$$\xi = \sqrt{-g_{55}}$$

The length of the loop is:

(7)

$$\int_0^{2\pi} \xi \, dx^5 \; \equiv 2\pi \, \xi$$

$\xi$ can be considered as the radius of the "tube-Universe".

In the following the equation (34.13) page 332 defines the action:

(8)

$$A \;=\; \int_C \sum_j p_j \, vol$$

The variational method gives the field equation, where the sum of the stress-energy tensors includes the gravitation and all other phenomena. The 5d field equation ([1], (41.59) page 405) is:

(9)

$$R_{jk} - \frac{1}{2} R \, g_{jk} + \Lambda \, g_{jk} = \chi \sum T_{jk} \quad with \quad j \; et \; k \in \{1,2,3,4,5\}$$

where $\Lambda$ is the cosmological constant and $\chi$ is Einstein's constant. As in the 4d approach we find:

(10)

$$\Lambda = -\frac{a}{2b} \qquad \xi = \frac{1}{2b}$$

The calculation provides the four-dimensional equations ([1], (41.61) p.405 and (41.62) p.406)

(11)

$$R_{\mu\nu} - \frac{1}{2} R g_{\mu\nu} + \Lambda g_{\mu\nu} \equiv \frac{\xi^2}{2} \left[ \frac{1}{2} g_{\mu\nu} \langle \mathcal{F}, \mathcal{F} \rangle - g^{\rho\sigma} \mathcal{F}_{\rho\mu} \mathcal{F}_{\sigma\nu} \right] + \chi \sum T_{\mu\nu}$$

$$+ \frac{\hat{\partial}_\mu \partial_\nu \xi - g_{\mu\nu} \square \xi}{\xi}$$

(12)

$$\text{div} \left[ \xi^3 \mathcal{F} \right]_\mu \equiv 2\chi\xi \sum \mathcal{J}_\mu$$

Writing the zero divergence condition J.M. Souriau gets the Maxwell equations ([1], p.407). Then he simplifies these equations, assuming the radius of the "tube-Universe" $\xi$ to be constant.

The last term of the second member of equation (11) becomes zero.

(13)

$$R_{\mu\nu} - \frac{1}{2} R g_{\mu\nu} + \Lambda g_{\mu\nu} \equiv \frac{\xi^2}{2} \left[ \frac{1}{2} g_{\mu\nu} \langle \mathcal{F}, \mathcal{F} \rangle - g^{\rho\sigma} \mathcal{F}_{\rho\mu} \mathcal{F}_{\sigma\nu} \right] + \chi \sum T_{\mu\nu}$$

## 2) Extension of the bigravity model to a five-dimensional geometry.

The equation (13) rules a Universe containing electrically charged particles. It is possible to build a first order solution, assuming the densities of electrical charge and electromagnetic energy are zero everywhere, which gives the simplified equation:

(14)

$$R_{\mu\nu} - \frac{1}{2} R g_{\mu\nu} + \Lambda g_{\mu\nu} \equiv \chi \sum T_{\mu\nu}$$

We add the hypothesis of a nil cosmological constant $\Lambda$.

(15)

$$R_{\mu\nu} - R g_{\mu\nu} = \sum T_{\mu\nu} \quad \text{with } \mu \text{ and } \nu \in \{1,2,3,4\}$$

which is the projection of the five-dimensional equation:

(16)

$$R_{jk} - R g_{jk} = \sum T_{jk} \quad \text{with } j \text{ and } k \in \{1,2,3,4,5\}$$

If the Universe is homogeneous and isotropic, the Robertson-Walker metric is a solution, with an extra term:

(17)

$$ds^2 = \left(d\,x^1\right)^2 - \left[R_{(x^1)}\right]^2 \ \frac{d\,u^2 + u^2(d\,\theta^2 + \sin^2\theta\, d\varphi^2\,)}{(1+\frac{k\,u^2}{4})^2} - dx^5$$

In order to link to the notations of former papers ([3], [4], [5]) we prefer to write the five variables:
(18)

$$\{\ x^\circ,\ x^1,\ x^2,\ x^3,\ \zeta\ \}\ \text{or}\ \{\ x^\circ,\ r,\ \theta,\ \varphi,\ \zeta\ \}$$

$x^\circ$ is the time-marker, $\{\ x^1,\ x^2,\ x^3\}$ or $\{\ r,\ \theta,\ \varphi\ \}$ the space markers and $\zeta$ the Kaluza coordinate, associated to the additional dimension. Introducing flatness ($k = 0$) we write the metric:
(19)

$$ds^2 = \left(d\,x^o\right)^2 - \left[R_{(x^{o1})}\right]^2 \ \left[\,d\,u^2 + u^2(d\,\theta^2 + \sin^2\theta\, d\varphi^2\,)\,\right] - dx^5$$

Introducing the function:
(20)

$$F = \left(\dot{x}^o\right)^2 - \left[R_{(x^{o1})}\right]^2 \ \left[\,\dot{u}^2 + u^2(\,\dot{\theta}^2 + \sin^2\theta\, \dot{\varphi}^2\,)\,\right] - \dot{x}^5$$

and the Lagrange equations:
(21)

$$\frac{d}{ds}\left(\frac{\partial F}{\partial \dot{x}^i}\right) = \frac{\partial F}{\partial x^i}$$

we get:
(22)

$$\ddot{\zeta}=0\ i.e \ :\varsigma=\alpha\,s+\beta\,(linear)$$

The introduction of an additional closed dimension does not change the results of our precedent papers. We can give the manifold M5 two five-dimensional metrics $g^+$ et $g^-$, solutions of the system of the following system of coupled field equations:
(24)

$$S^+_{\ jk} = R^+_{\ jk}\ -\frac{1}{2}\ R^+ g^+_{\ jk} = \chi\ (\ T^+_{\ jk} - T^-_{\ jk}\ )$$

$$S^-_{\ jk} = R^-_{\ jk}\ -\frac{1}{2}\ R^- g^-_{\ jk} = \chi\ (\ T^-_{\ jk} - T^+_{\ jk}\ )$$

The two coupled metrics:

(25)

$$ds^{+2} = \left(d\,x^o\right)^2 - \left[R^+_{(x^{o1})}\right]^2 \ \left[\,d\,u^2 + u^2(d\,\theta^2 + \sin^2\theta\, d\varphi^2\,)\,\right] - dx^5$$

(26)

$$ds^{-2} = \left(dx^o\right)^2 - \left[R^-_{(x^{o1})}\right]^2 \left[du^2 + u^2(d\theta^2 + \sin^2\theta\, d\varphi^2)\right] - dx^5$$

are solutions of the system and give the evolution of the two space scale factors versus the time marker $x^\circ$

(27)

$$\frac{d^2R^+}{dx^{o2}} = -\frac{1}{(R^+)^2}\left(1 - \frac{(R^+)^3}{(R^-)^3}\right)$$

$$\frac{d^2R^-}{dx^{o2}} = -\frac{1}{(R^-)^2}\left(1 - \frac{(R^-)^3}{(R^+)^3}\right)$$

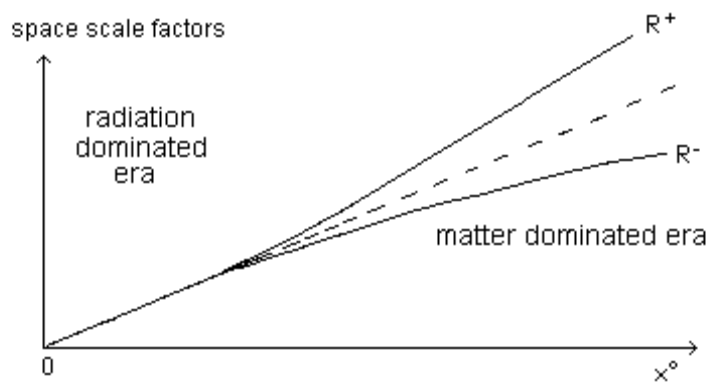

Fig.1: **Evolution of the space scale factors with respect to the time marker x°**

As presented in reference [2] this explains the acceleration of the expansion of our Universe, supposed to be described by the metric $g^+$ (positive energy matter).

In the next paper [3] we recall that the Roberston-Walker metric is based on the assumptions of isotropy and homogeneity, which does not fit the observational data. The expansion occurs in large portion of "empty" space (in fact filled by primeval photons). Elsewhere, space does not expand. This corresponds to a more refined metric solution, that we don't hold right now. Introducing a 2d didactic model, the Universe looks like a diamond, with blunt summits.

In the precedent paper we gave a very simplified 2d toy model: a cube with eight blunt corners. Each of these corners is an eighth of a sphere, that does not expand. These blunt summits are linked by Euclidean surfaces: quarters of a cylinder and flat plates, which expand. On figure 2 the O(2) symmetry is broken in step 2.

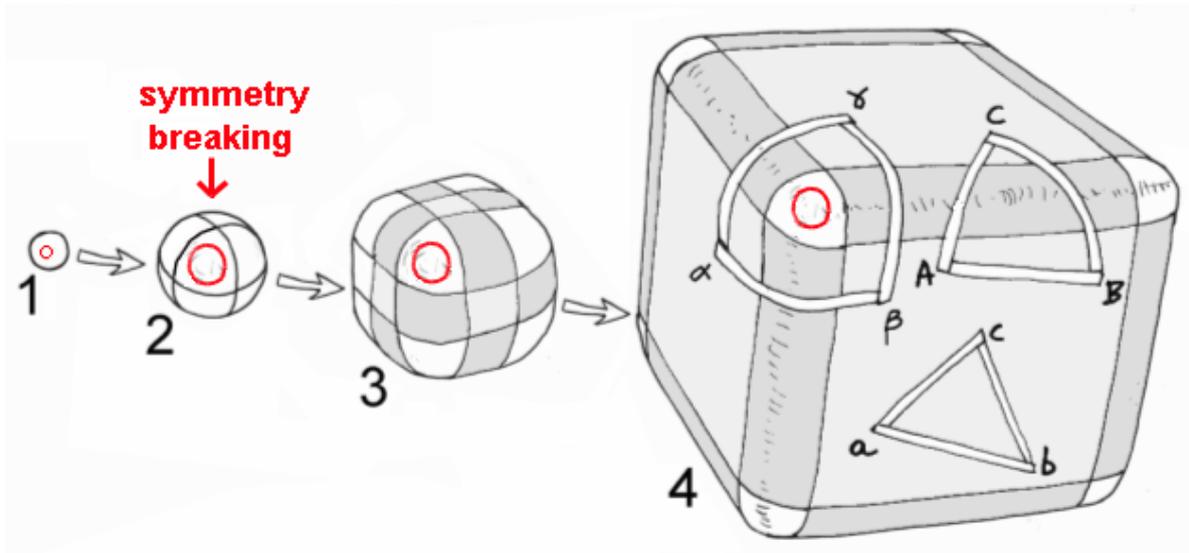

Fig.2: **2d didactic model with symmetry breaking phenomenon**

We have assumed that this symmetry breaking occurred at a undefined moment, early located in the radiation dominated era.

- After this symmetry breaking the constants of physics behave like absolute constants.

- Before, they are involved, with space and time scale factors, in a generalized gauge process.

## 4) Universe with finite lifetime matter

We don't know if matter has a finite lifetime, or not. Attempts has been done to evidence that the proton could have a finite lifetime. But this particle did not cooperate. This does not shows that the proton has an infinite lifetime, only that our experimental apparatus did not succeed to give the answer to that question.

We can assume that all massive particles, with positive or negative energy, have finite lifetimes. Then, after they have all decayed the Universe would recover the O(3) symmetry and would be ruled again by the generalized gauge process, involving space and time scale factors and variable constants. As a consequence, the speed of light would tend to zero:

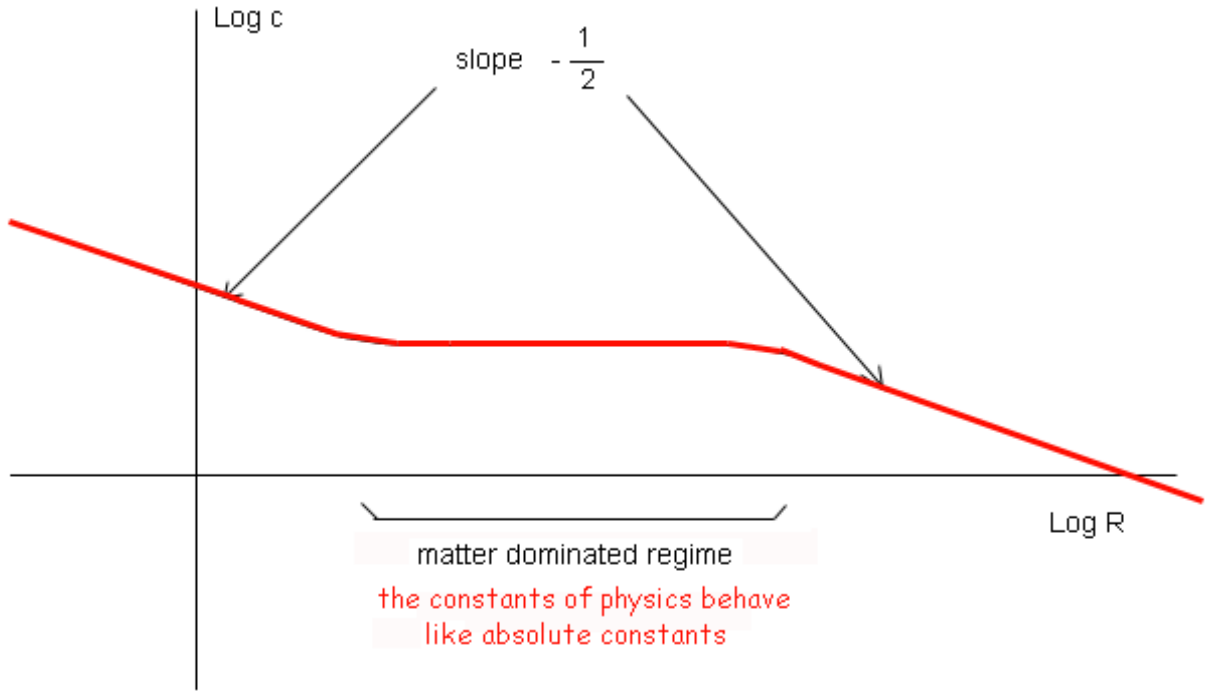

Fig.3: **Evolution of the speed of the light versus the space scale factor R.**

This corresponds to the following 2d didactic image:

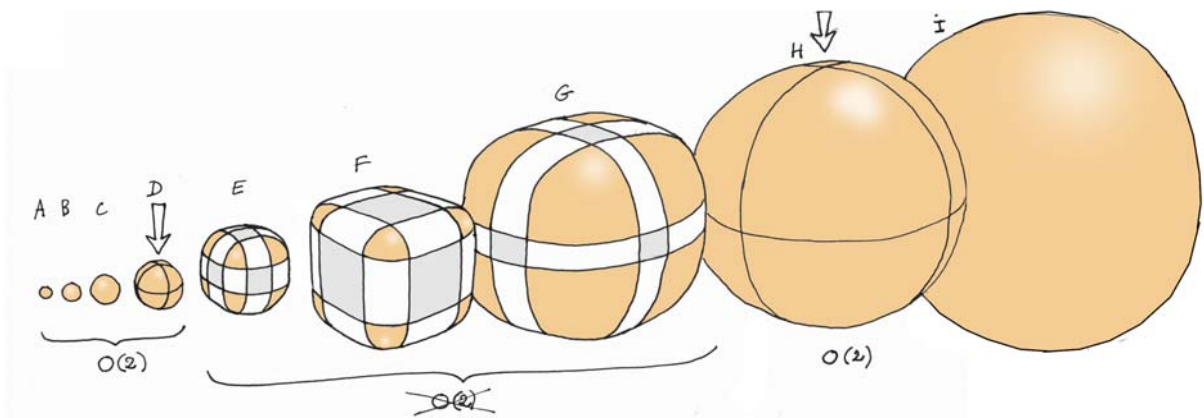

Fig.4: **2d didactic image of this new model**

The images A, B, C, D of figure 4 correspond to a generalized gauge process evolution. The symmetry breaking ( here: O(2) ) occurs in D. Between D and H we have a mixed geometry. In H the O(2) symmetry is recovered and the generalized gauge phenomenon rules the evolution.

## 5) **Topological structure**

The following ideas were introduced in 1994 in reference [4]. In more recent papers we described the Universe as a manifold equipped with two metrics $g^+$ and $g^-$. This description is better accepted by mathematicians although it becomes then impossible to get any mental representation of the object. How to imagine that between two given comoving points A and B we can draw two different geodesics and measure two different distances ?

Anyway it is possible to consider that the manifold, coupled to the metric $g^+$ forms a first hypersurface, whose points can be called $M^+$ and that the same manifold, coupled to the metric $g^-$ forms a second hypersurface whose points can be called $M^-$. The (naked) manifold provides a point to point mapping between the two hypersurfaces.

In addition, in the paper [4] the Universe, supposed to be compact, was assumed to be the two-fold cover of a projective space P4.

As a good didactic image let us start from a closed 2d space-time, figured by a sphere. The Big Bang is one of its poles. The Big Crunch is the second one. The meridians form the world-lines system and the parallel the closed space at successive epochs. The equator of the sphere represents the maximum space extension.

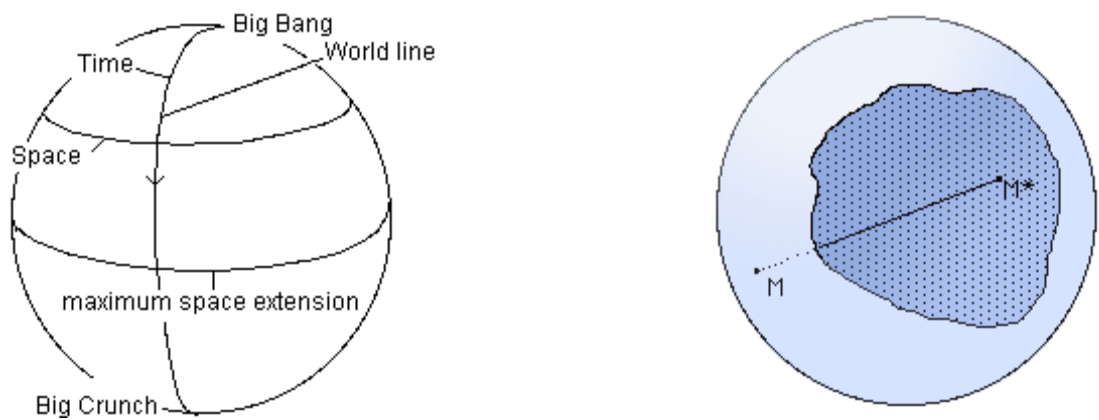

Fig.5: **2d spherical space-time**

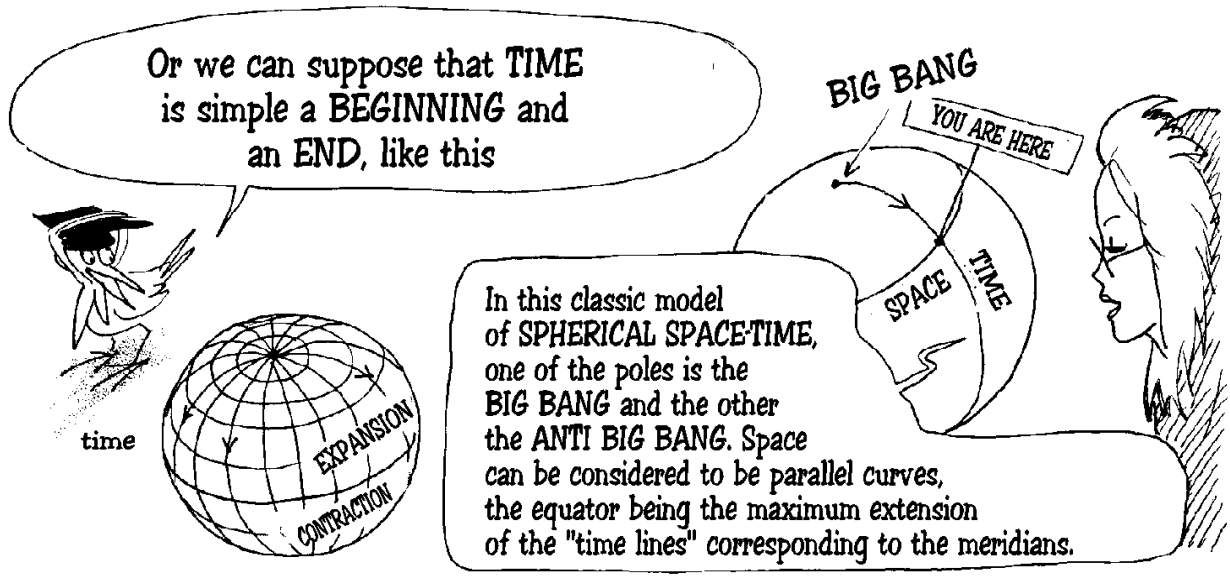

Suppose M and M* are a couple of antipodal points. We know that we can make any point of this sphere to coincide with its antipode, the result being the well-known Boy surface, an immersion of the projective space P2 in R3.

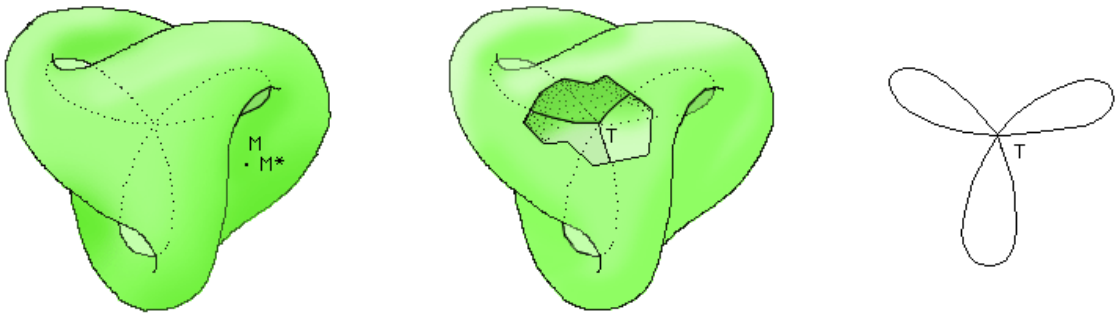

Fig.6: **The Boy an its self-intersection curve, with its triple point T.**

To become familiar to such stuff, read my (free) comic book " Topo the World", 1985 [8].

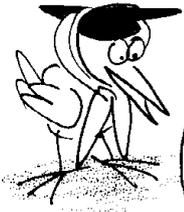
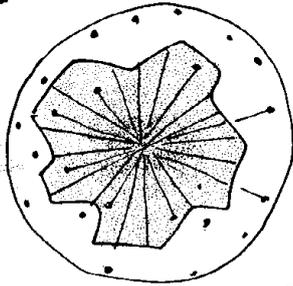

We begin by joining each point of the sphere to its antipode using strings soaked in SHRINKASOL

These strings contract to the point where they have zero length, while the surface of the sphere remains constant. We bring each point into CONJUNCTION with its ANTIPODAL.

But as you'll see all that in another album, dedicated to turning a SPHERE inside out. In the meantime, the series of images in the 'filmstrip' G show how the EQUATOR of the SPHERE folds in on itself, becoming the EQUATOR of the BOY. The NORTH pole then, obviously, sticks itself next to the SOUTH pole.

*The Management*

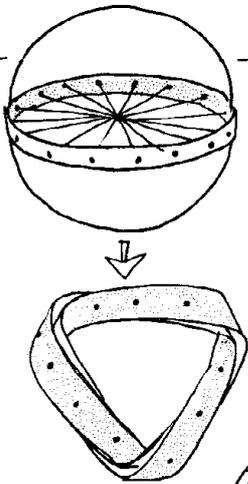
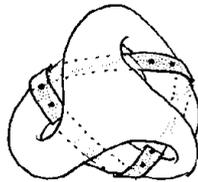
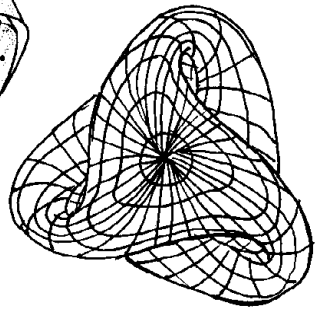
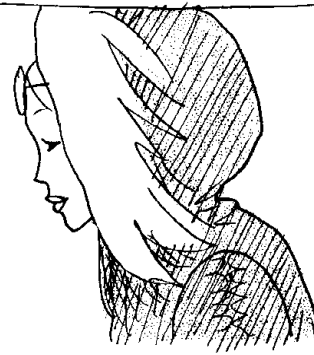

All the meridians and parallels of the sphere cover each other.

The main point to keep in mind is that the self-intersection structure and the triple point are not intrinsic geometric attributes of the object but simple consequences of the peculiar space in which the object is placed (R3). In figure 7 we show the vicinity of a meridian circle. Two opposite arrows of time emerge from the "north pole", the Big Bang. They focus on the antipodal pole, the Big Crunch.

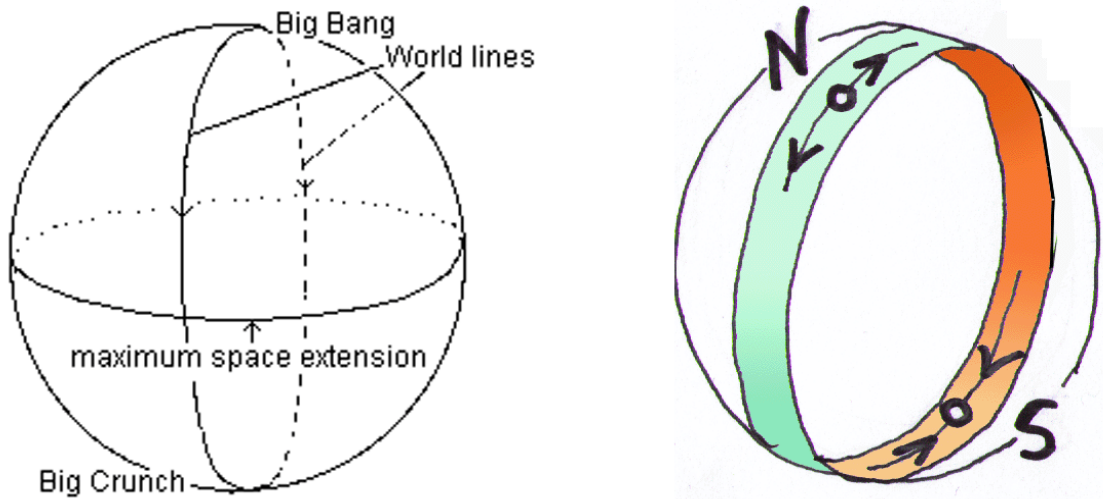

Fig.7: **World lines on a 2d spherical closed space-time**

The figure 8 shows this strip, vicinity of the meridian, glued on itself as the two fold cover of a one half turn single sided Moebius strip. The two poles coincide in Φ.

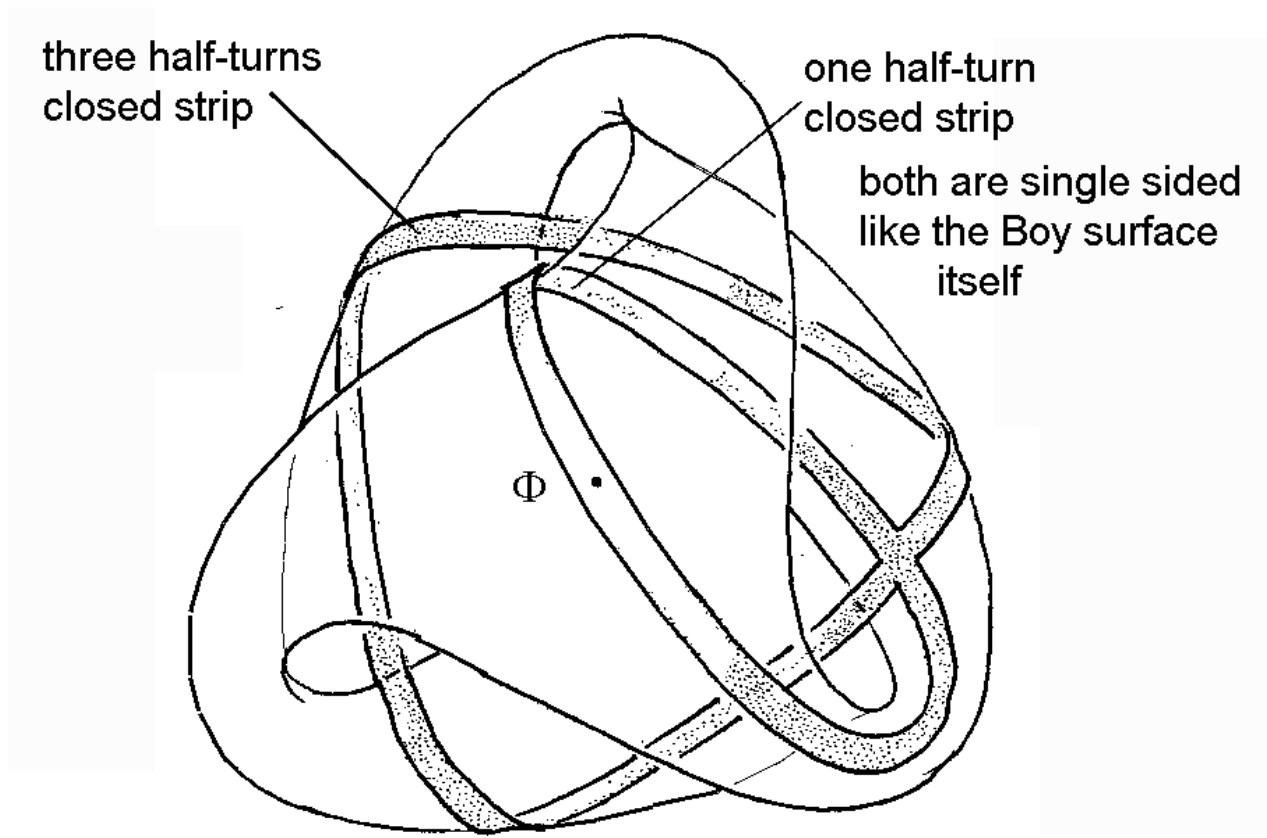

Fig.8: **Vicinities of meridian and equator lines arranged onto a Boy surface**

Antipodal regions of the S2 sphere become adjacent. The arrows of time of adjacent regions become everywhere anti-parallel. Following, the vicinity of a world line, glued on itself along a one half-turn closed strip (classical single sided Moebius strip) .

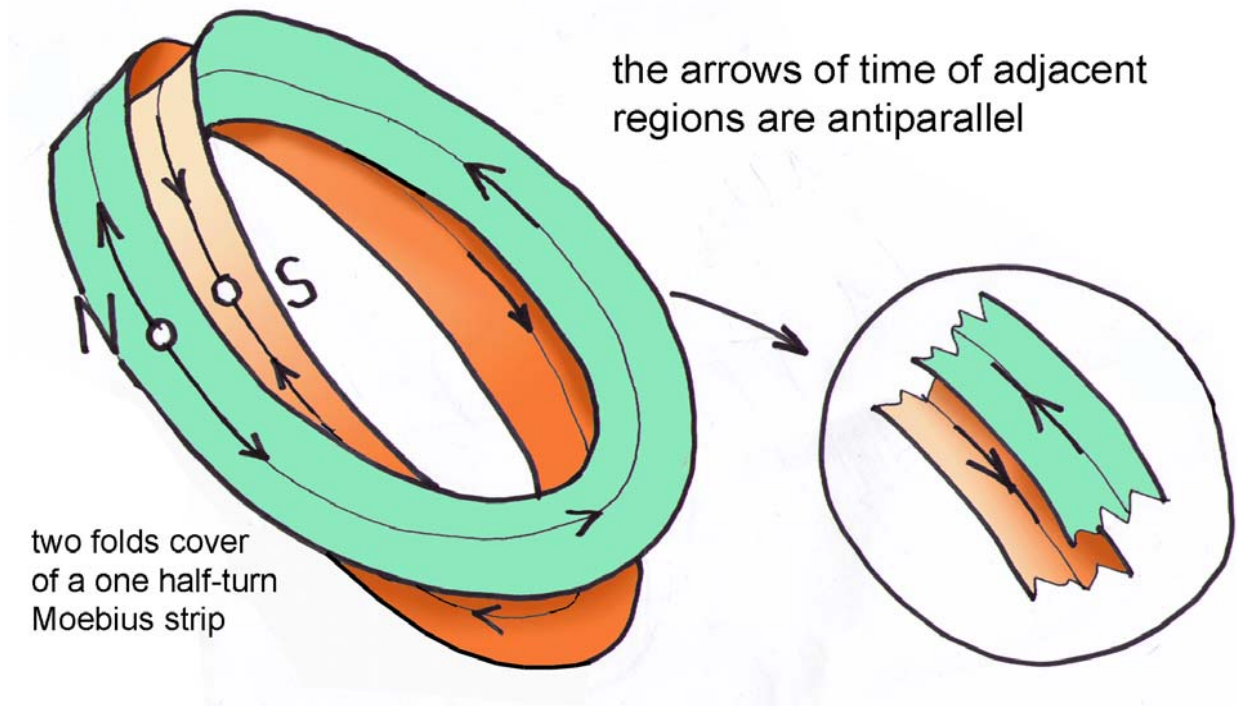

Same thing for the vicinity of the maximum space extension (equator of the S2 sphere) glued on itself along a single sided three half turns Moebius strip:

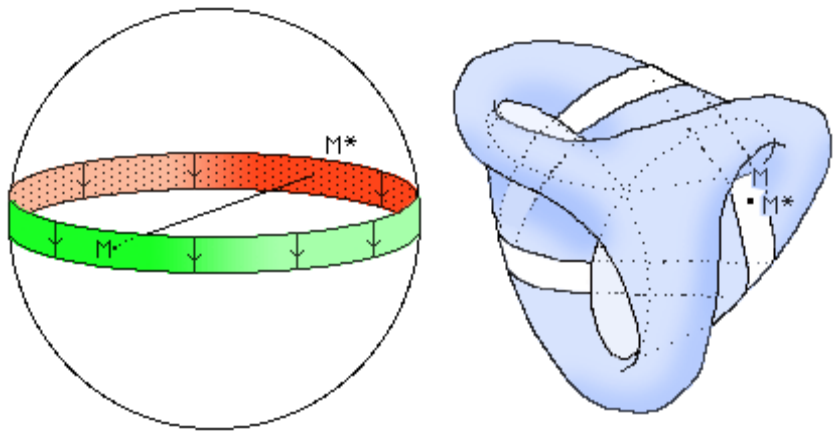

Fig.9: **Equator of the spherical space-time S2. Arrows of time**

On the right image of figure 8 we find the three half-turns Moebius strip. The figure 10 shows the inversion of the arrow of time and the enantiomorphical relationship (P-symmetry).

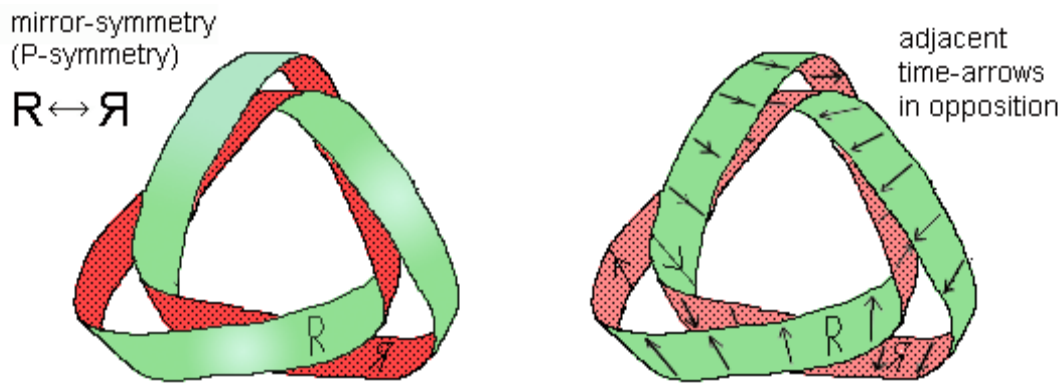

Fig. 10: **The equator of the spherical 2d space time, shaped as the two fold cover of a three half-turns Moebius strip showing both the apparent inversion of the arrow of time and the enantiomorphy (mirror-symmetry) of associated portions of the surface.**

By the way, notice that the adjacent regions are enantiomorphic ( P-symmetric).

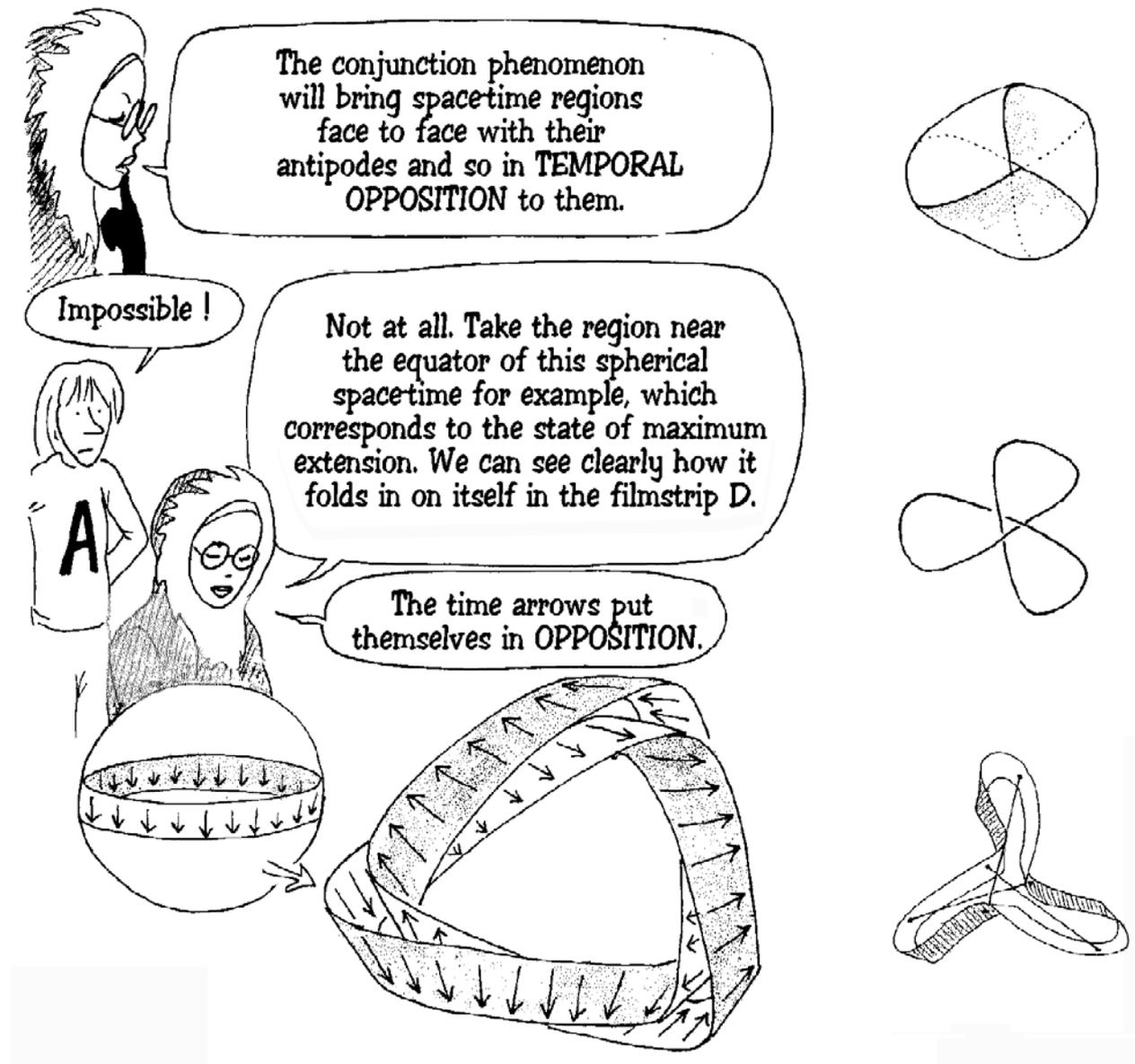

The two poles of the sphere coincide. In 1967 [7] Andrei Sakharov imagined to figure the Universe as a set of two "twin universes", with opposite arrows of time, joined by a singularity Φ. Following, the 2d didactic image of this schema:

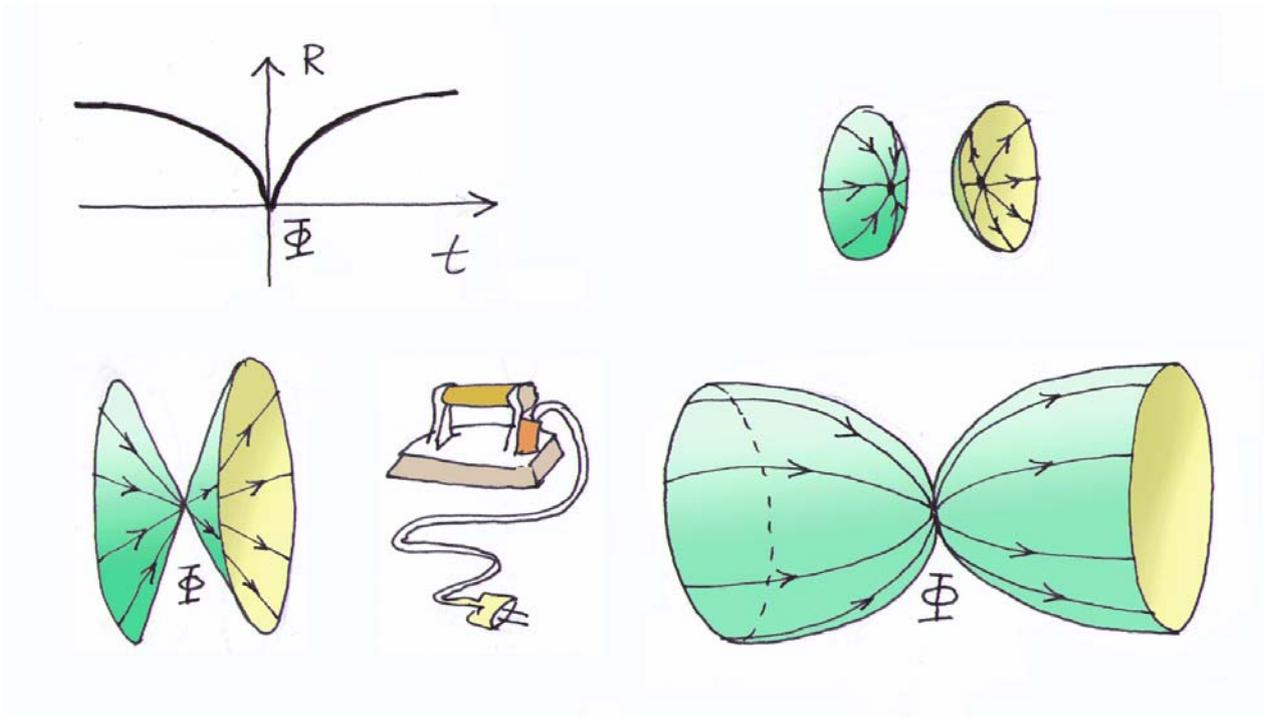

Fig.11: **Didactic 2d image of Sakharov's model
with twin Universes linked by a singularity. Φ**

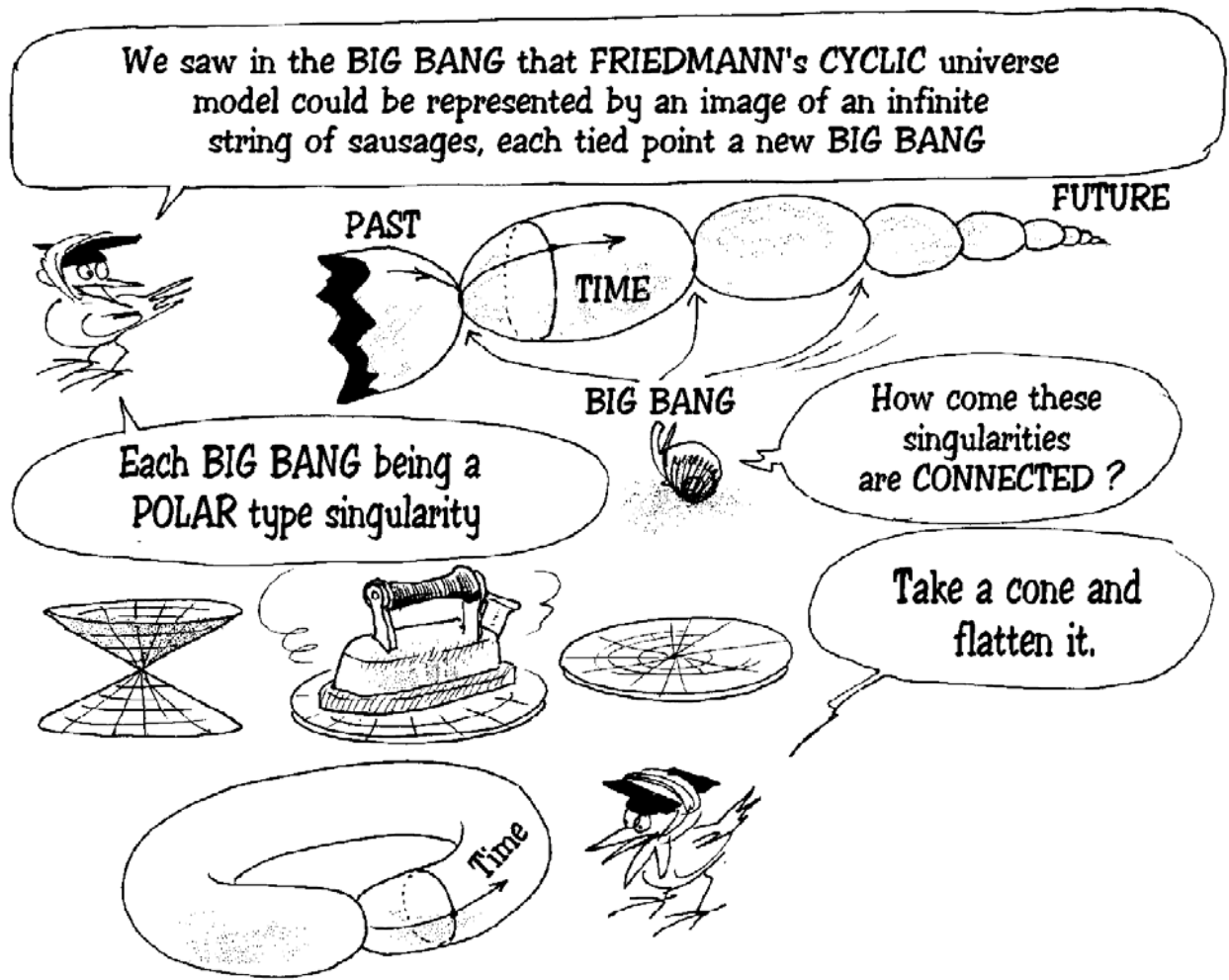

Starting from two cones joined by their summits and using a flatiron we can ensure the continuity of the geodesics, through the Sakharov singularity Φ.

At this level this 2d geometrical toy model is an image of a Universe composed by two coupled folds with opposite arrows of time. This can be extended to a closed 4d structure. Then the Universe becomes the two folds cover of a projective space P4.

P2 and P4 have the same Euler-Poincaré characteristic, whose value is unity. They can be mapped with a single pole. The maximum space extension (the "equator" of the hypersurface) is a S3 sphere. As suggested in 1994 the inversion of time in Sakharov's model can be derived on topological grounds, considered as the consequence of a two-fold cover representation.

Notice that this geometrical configuration is equivalent to a bimetric description, each "adjacent fold portions of Universe" being defined by its metric. According to [6] the inversion of time goes with the inversion of energy and mass.

6) **Back to the symmetry breaking and symmetry recovering.**

The space-time 2d image, corresponding to figure 4 is the following. In a distant past, early in the radiation-dominated era a symmetry breaking occurs. According to this image, there is some "origin in time" but no end in the future. The white areas represent the region where the O(2) symmetry holds.

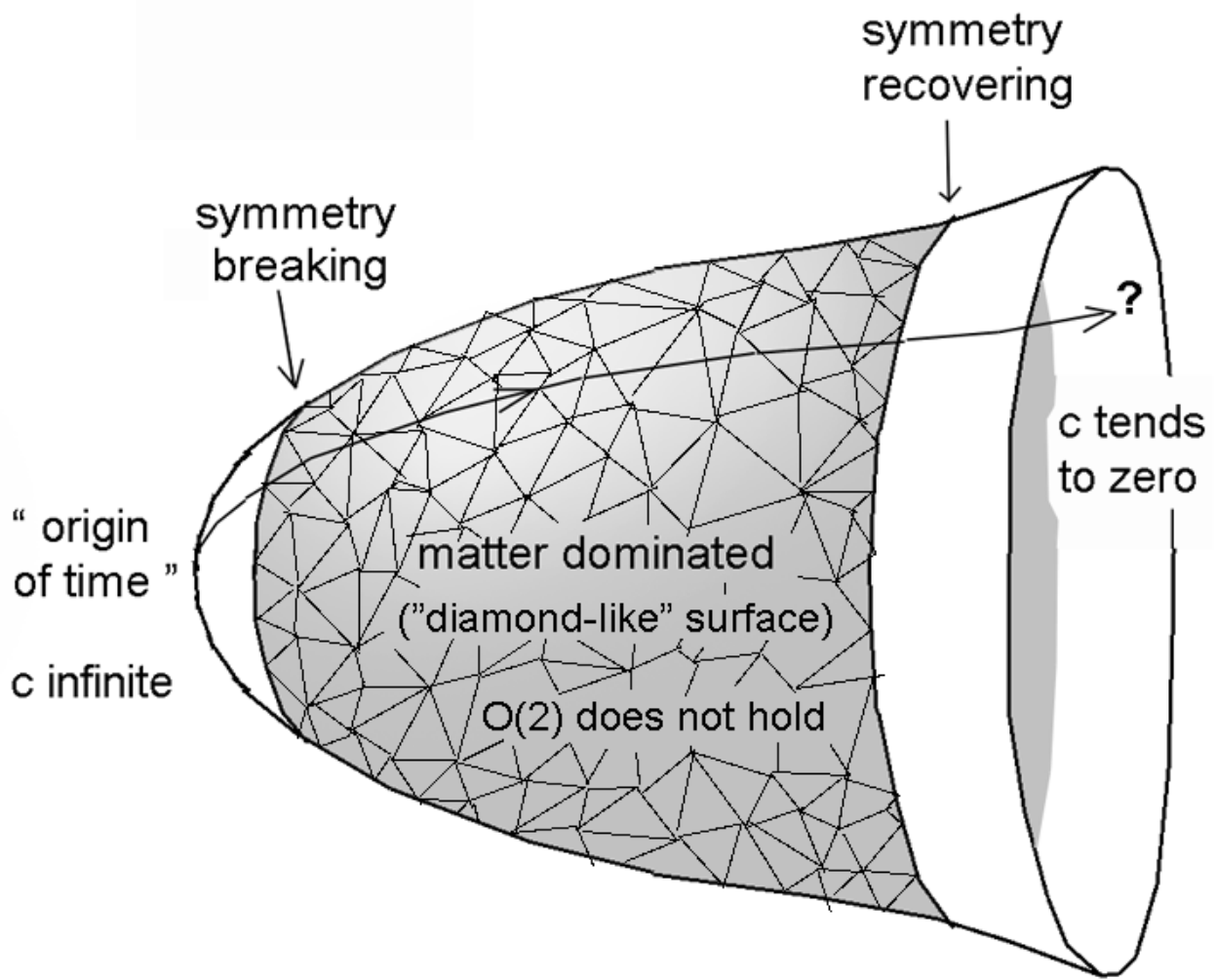

Fig.12: **2d space time image with symmetry breaking and recovering**

The diamond-like portion of the surface illustrates that, in such place the O(2) symmetry does not hold. Il we choose a fully closed universe (spherical space-time) we get the figure 12.

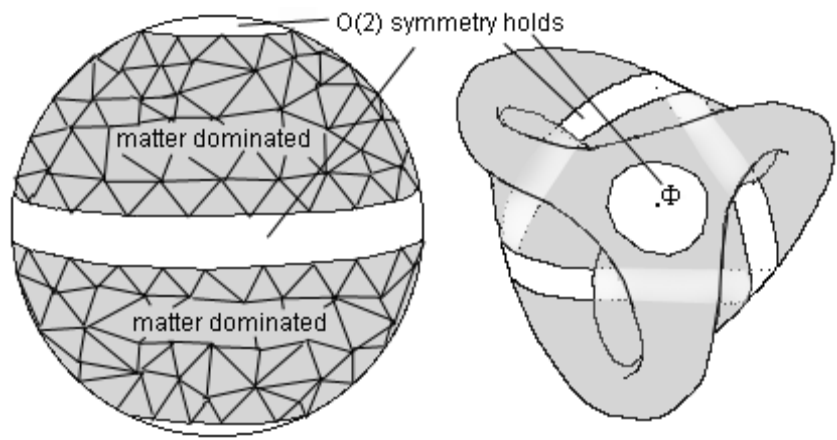

Fig.13**: Closed 2d Universe with symmetry breaking and recovering.
In white areas the O(2) symmetry holds**

The diamond-like portions of surface suggest a region where the O(2) symmetry is broken. On the right, the Greek letter Φ is the Sakharov singularity. The figure 13 is a 2d didactic representation of Sakharov's Φ singularity.

## 7) **A Universe without "initial singularity"**

Inspired by an egg-timer we can modify the precedent figure in order to eliminate the singularity, replaced by a narrow throat. In addition, if we assume that the Universe is closed, the corresponding 2d didactic model is a torus T2:

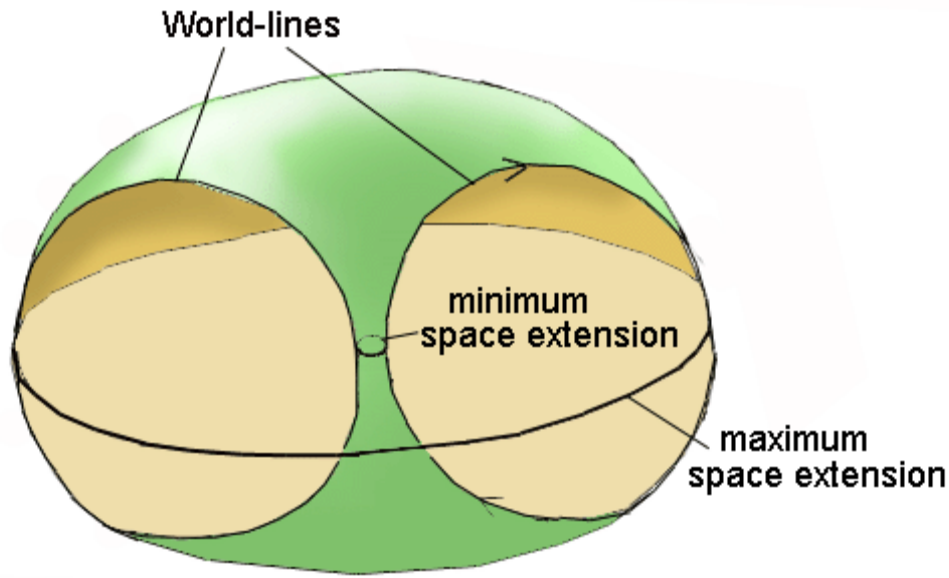

Fig.14: **2d torus world**

Then we can shape this T2 torus as the two-fold cover of a Klein bottle K2.

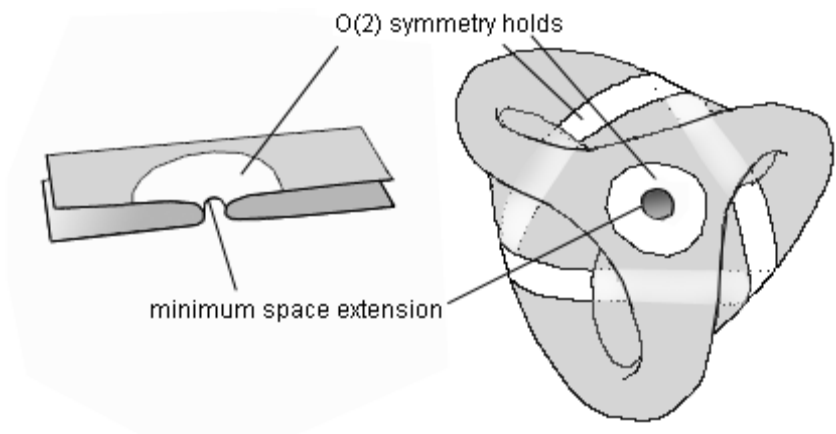

Fig15: **Two-folds cover of a Klein bottle K2**

This is indeed a torus, for the small tube gives the Euler-Poincaré characteristic a nil value;
Of course the following sketch will be more familiar to the reader.

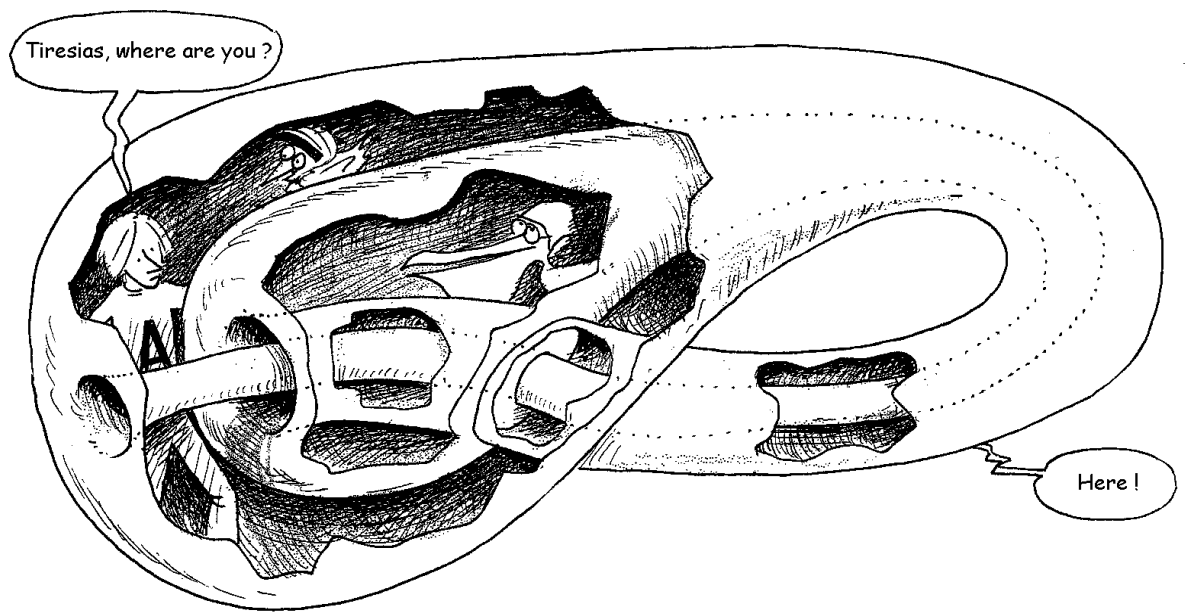

Fig.16: **More familiar view of the 2-folds cover of a Klein's bottle K2**

## 8) Conclusion: a new geometric framework for the Universe.

In reference [2] we presented a bimetric model of the Universe, which explained the acceleration of the expansion, at large redshifts. Then in [3] we added symmetry breaking and variable constants technique, which explained the homogeneity of the primeval Universe. Now, back to an older work [4] we add some topological features. The Universe is described as the two-fold cover of a projective space P4. The analysis of the world-lines evidences the inversion of the arrow of time. Thanks to the work of J.M. Souriau [6] this inversion of time justifies the inversion of masses and energies, as described by the second metric $g^{-}$ Adding a closed fifth dimension we deal with electrically charged particles.

Let's sum up our assumptions:

- The universe is closed in all its dimensions, has a maximum space extension
- There is no initial singularity, so that it owns a minimum space extension, small size, which replaces the Big Bang singular structure.
- When going backwards in time, towards the minimum space extension, a symmetry breaking occurs, during the radiation-dominated era. Then, as described in [3] this first era goes with variable constants, linked to space and time scale factor through a generalized gauge process. Close to the minimum space extension the value of the speed of light tends to a very large value, "infinite".

- We assume that matter, with positive or negative mass has a finite lifetime, even the latter is very large. Then we assume that the O(3) symmetry is recovered in a distant future. As a consequence the generalized gauge process restarts and rules the cosmic evolution. The constants of physics vary. In particular the speed of light tends to an almost nil value.
- The dynamics of the Universe corresponds to a bimetric description, going with a system of two coupled field equations [3]. It contains two kinds of species, with positive or negative mass or energy.
- This can be derived from geometrical grounds. The inversion of the arrow of time of one species comes from the geometrical structure of the Universe, considered as the two-fold cover of a projective, which brings both T and P-symmetry. The inversion of the mass and energy goes with the inversion of the time (T-symmetry). Electric charge can be given to the relativistic mass-points, considering the Universe as a five-dimensional bundle. As a first order solution, the Kaluza metric fits the five dimensional field extension, as introduced by J.M. Souriau in 1964.

Choosing the description based on two coupled five-dimensional geometrical structures, inspired by the vision of Andrei Sakharov [7]: twin universe model, we may think about the geometric structure of the two boundary 4d-universes, linking the two, when the velocity of the light is "almost infinite" or "almost nil". Among the five dimensions, one must disappear. The five dimensions are:

- Space ( x , y , z )
- Time t
- Fifth dimension $\zeta$

P-symmetry holds in our fold of the Universe (in our positive energy world). The polarization of photons is for an example a proof that such symmetry can be found in nature.

$\zeta$-symmetry holds. As shown in [1] this corresponds to charge conjugation and matter-antimatter duality. We find antimatter in our positive energy world, so that $\zeta$-symmetry is evidenced in natural phenomena.

We don't find negative energy particles. The "second world", with negative mass and energy corresponds to T-symmetry. As a conclusion, in the two metrics associated to the boundary universe, time disappears and the signature becomes Euclidean:

( - - - - )

The length, in the two boundary Universes (one corresponding to an "infinite" value of the speed of light and the second to a "nil" value of this velocity), is an imaginary quantity. The figure 17 is a didactic image showing the geometric structure of the "Big Bang" according to this model:

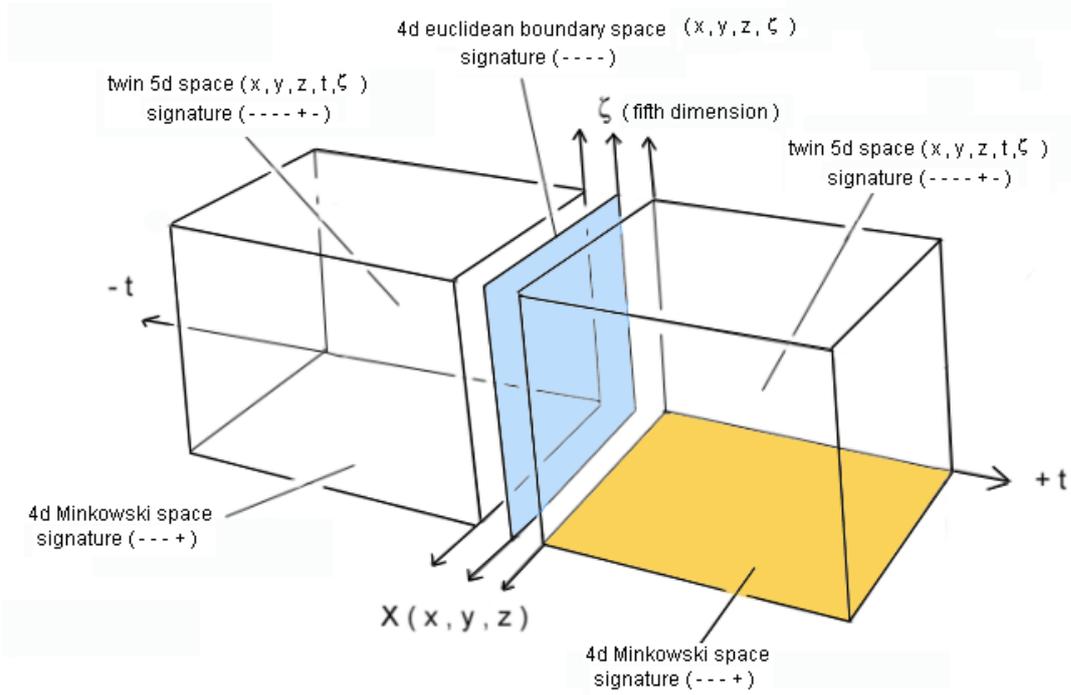

Fig. 17: **Two portions of five-dimensional "twin spaces"
linked by an Euclidean four dimensional boundary space.**